\documentclass[letterpaper]{article} 

\pdfoutput=1
\usepackage{comment}

\usepackage{aaai2026}  
\usepackage{times}  
\usepackage{helvet}  
\usepackage{courier}  
\usepackage[hyphens]{url}  
\usepackage{graphicx} 
\usepackage{natbib}  
\usepackage{caption} 
\usepackage{algorithm}
\usepackage{algorithmic}

\usepackage{amsmath,amsfonts}

\usepackage{array}
\usepackage[caption=false,font=normalsize,labelfont=sf,textfont=sf]{subfig}
\usepackage{textcomp}

\usepackage{verbatim}

\usepackage{cite}

\usepackage{newfloat}
\usepackage{listings}
\DeclareCaptionStyle{ruled}{labelfont=normalfont,labelsep=colon,strut=off} 
\floatstyle{ruled}
\newfloat{listing}{tb}{lst}{}
\floatname{listing}{Listing}
\title{MyGO: Make your Goals Obvious, Avoiding Semantic Confusion in Prostate Cancer Lesion Region Segmentation}
\author{
    Zhengcheng Lin\textsuperscript{\rm 1},
    Zuobin Ying\textsuperscript{\rm 1},
    Zhenyu Li\textsuperscript{\rm 2},
    Zhenyu Liu\textsuperscript{\rm 3}
    Jian Lu\textsuperscript{\rm 4}
    Weiping Ding\textsuperscript{\rm 5}
}
\affiliations{
    
    \textsuperscript{\rm 1}City University of Macau\\
    \textsuperscript{\rm 2}Shandong University\\
    \textsuperscript{\rm 3}Chinese Academy of Sciences\\
    \textsuperscript{\rm 4}Peking University\\
    \textsuperscript{\rm 5}Nantong University\\

}

\usepackage{bibentry}

\begin{document}
	
	\maketitle
	
	\begin{abstract}
		Early diagnosis and accurate identification of lesion location and progression in prostate cancer (PCa) are critical for assisting clinicians in formulating effective treatment strategies. However, due to the high semantic homogeneity between lesion and non-lesion areas, existing medical image segmentation methods often struggle to accurately comprehend lesion semantics, resulting in the problem of semantic confusion.
		To address this challenge, we propose a novel Pixel Anchor Module, which guides the model to discover a sparse set of feature anchors that serve to capture and interpret global contextual information. This mechanism enhances the model’s nonlinear representation capacity and improves segmentation accuracy within lesion regions. Moreover, we design a self-attention-based Top\_$k$ selection strategy to further refine the identification of these feature anchors, and incorporate a focal loss function to mitigate class imbalance, thereby facilitating more precise semantic interpretation across diverse regions.
		Our method achieves state-of-the-art performance on the PI-CAI dataset, demonstrating 69.73\% IoU and 74.32\% Dice scores, and significantly improving prostate cancer lesion detection.
	\end{abstract}
	
	\begin{links}
		\link{Code}{https://github.com/LZC0402/MyGO}
		\link{Datasets}{https://pi-cai.grand-challenge.org/}
	\end{links}
	
	\section{Introduction}
	Prostate cancer represents a major urological disease affecting middle aged and elderly men globally \cite{rawla2019epidemiology}. According to GLOBOCAN 2022 \cite{bray2024global}, there were 1,466,680 new cases and 396,792 deaths worldwide in 2022. In 2025, an estimated 313,780 new cases of prostate cancer will occur in the United States and approximately 35,770 men will die from prostate cancer, making it the most common cancer in men and accounting for approximately 30\% of all male cancers \cite{siegel2025cancer}. Currently, transrectal ultrasound guided biopsy is the mainstream approach for PCa screening in clinical practice; however, it may lead to multiple complications and often requires repeated procedures due to sampling errors. An alternative strategy involves magnetic resonance imaging (MRI), including T2-weighted imaging (T2WI), diffusion weighted imaging (DWI), and apparent diffusion coefficient (ADC) maps, to detect suspicious lesions \cite{tanimoto2007prostate}. Once a lesion is detected, accurate segmentation from surrounding tissues is critical for subsequent cancer grading and treatment planning.
	
	\begin{figure}[t]
		\centering
		\includegraphics[width= 0.875\columnwidth]{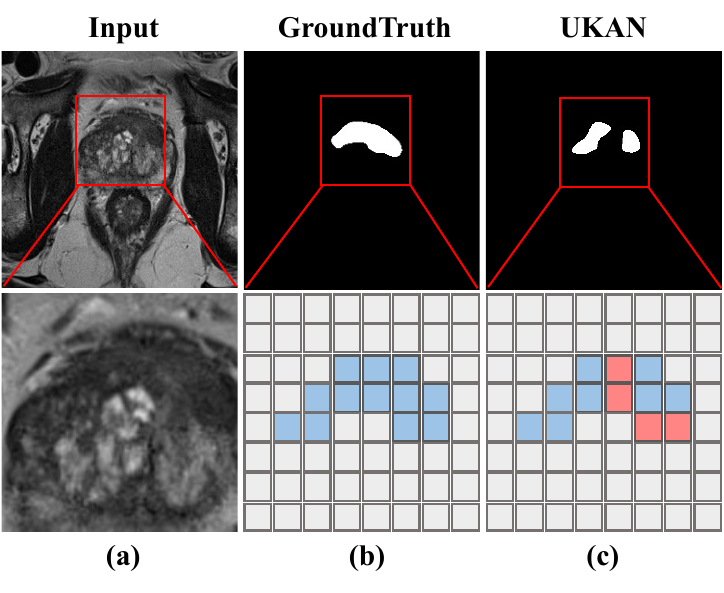} 
		\caption{(a) The input MRI image and its local zoom-in view; (b) The ground truth segmentation and its corresponding magnified region; (c) The output of U-KAN applied to the input image. The red regions indicate false negative predictions resulting from semantic confusion, where the model fails to recognize true lesion areas.}
		\label{fig_1}
	\end{figure}
	
	With the rapid advancement of computer science, the use of computer-assisted techniques for disease detection and diagnosis in medical imaging has significantly improved. Although various deep learning-based methods for tumor segmentation in MR images have been proposed in recent years, existing approaches still face a key limitation: difficulty in capturing the complete semantic features of all lesion regions. This is largely due to the high semantic similarity between tumor tissues and adjacent non-tumorous structures in MR images, which challenges accurate lesion characterization and leads to semantic confusion, resulting in increased false positives or false negatives. Consequently, some lesion regions may be missed, as illustrated in \textbf{Figure 1}, potentially compromising cancer grading and clinical decision-making.

	Recent advances have led to a growing body of work on computer-aided diagnosis of prostate cancer using MRI-based segmentation \cite{ellmann2020computer,dinh2018characterization}. The development of deep learning techniques has further demonstrated new opportunities in leveraging MRI segmentation for PCa detection \cite{wildeboer2020artificial}, with several studies \cite{duran2022prostattention,le2017automated} showcasing their superior performance in clinical applications. Currently, these two-dimensional convolutional neural networks (2D CNNs) remain the dominant approach for tumor segmentation. However, under conditions where semantic features of tumor shadows closely resemble those of surrounding normal tissues, these models often struggle to capture the complete semantic characteristics of all lesion regions. This limitation stems from the high homogeneity of semantic features, which renders the models less sensitive to subtle boundary features of the tumor shadows.
	To mitigate semantic confusion and enhance the efficiency of feature discrimination between lesion and surrounding tissues, we propose a novel pixel-pivot module inspired by recent work \cite{park2023self}. This module selects a representative pixel within a region as an anchor to guide regional feature aggregation, significantly reducing the computational overhead for semantic representation. We adopt U-KAN \cite{li2025u} as the backbone due to its strong nonlinear modeling capacity, which complements the pixel-pivot mechanism in capturing salient semantic anchors. The module enables the network to extract and generalize global features based on a sparse set of pivotal anchors, self-attention mechanism assigns region-specific weights, guiding anchor selection constrained by Pixel-wise Cross-entropy and Focal Loss, thereby improving semantic differentiation between lesion and non-lesion areas. As a result, our method exhibits superior performance in distinguishing tumor regions from surrounding tissues, particularly under conditions where semantic homogeneity leads to blurred lesion boundaries.
	
	Our contributions are as follows:
	
	\begin{itemize}
		
		\item We propose a segmentation strategy tailored for prostate cancer lesion analysis, which employs inter-correlated feature anchors extracted from the feature map to encode global contextual semantics. This strategy substantially enhances the model’s ability to identify small-scale lesions and disambiguate visually similar regions, thereby mitigating semantic confusion induced by high inter-region homogeneity.
		
		\item We propose a module which guided by a self-attention Top\_$k$ selection, we call it as Pixel Anchor Module. The module that enables adaptive extraction of representative anchors across feature maps. In conjunction, we design a novel Pixel Anchor Module to semantically decode these anchors by leveraging surrounding contextual dependencies. This module significantly boosts representational efficiency and fosters improved semantic differentiation during the feature refinement process.

		\item Extensive experiments conducted on the PI-CAI benchmark demonstrate the superiority of our approach, achieving an IoU of 69.73\% and a Dice coefficient of 74.32\%. Our method outperforms existing medical image segmentation methods, establishing state-of-the-art results in multiple evaluation metrics.
		
	\end{itemize}

	\section{Related Works}
	
	\subsection{U-Based Methods}
	
	
	Before 2018, most medical image segmentation methods relied on convolutional neural networks (CNNs) \cite{10643318}, especially U-Net \cite{ronneberger2015u} and its variants\cite{he2016deep}. The introduction of residual networks \cite{he2016deep} brought major improvements, leading to models like \cite{drozdzal2016importance} and V-Net \cite{milletari2016v} in 2016, which were successfully applied to medical tasks. In 2017, the emergence of attention mechanisms \cite{vaswani2017attention} led to models such as Attention U-Net\cite{oktay2018attention} and its upgrade, Attention U-Net++ \cite{li2020attention}, in 2020. With the development of Transformer architectures, hybrid models like TransUNet \cite{chen2021transunet}, Swin-UNet \cite{cao2022swin}, and UCTransNet \cite{wang2022uctransnet} combined CNNs with Transformers, achieving strong results. More recently, advanced approaches such as RollingUNet, DCF-Net, U-Mamba, and U-KAN \cite{liu2024rolling, 10650540, ma2024u, li2025u} have pushed performance even further, setting new state-of-the-art benchmarks.
	
	\subsection{Lesion Detection and Segmentation}
	
	Prior to 2018, most approaches for detecting and segmenting prostate cancer lesions were based on convolutional neural networks (CNNs). Such as, Msak-RCNN \cite{He_2017_ICCV}was employed for lesion detection in prostate MRI scans. With advancements in machine learning, these methods \cite{ellmann2020computer, dinh2018characterization} became representative computer-aided detection techniques for identifying prostate cancer lesions.
	As deep learning techniques matured, segmentation of prostate MRI scans containing suspected lesions emerged as a viable approach for prostate cancer diagnosis \cite{wildeboer2020artificial}; In the past five years, methods for prostate cancer detection and grading using bi-parametric MRI have been proposed \cite{9090311, mehralivand2022deep}. In 2024, a fully automated deep learning model for prostate cancer detection via MRI was introduced \cite{cai2024fully}. Research efforts \cite{arif2020clinically, duran2022prostattention, aldoj2020semi, le2017automated, zhong2019deep} further demonstrated the superiority of deep learning methods in prostate cancer lesion detection and segmentation.
	
	\begin{figure*}[!t]
		\centering
		\includegraphics[width=0.9725\textwidth]{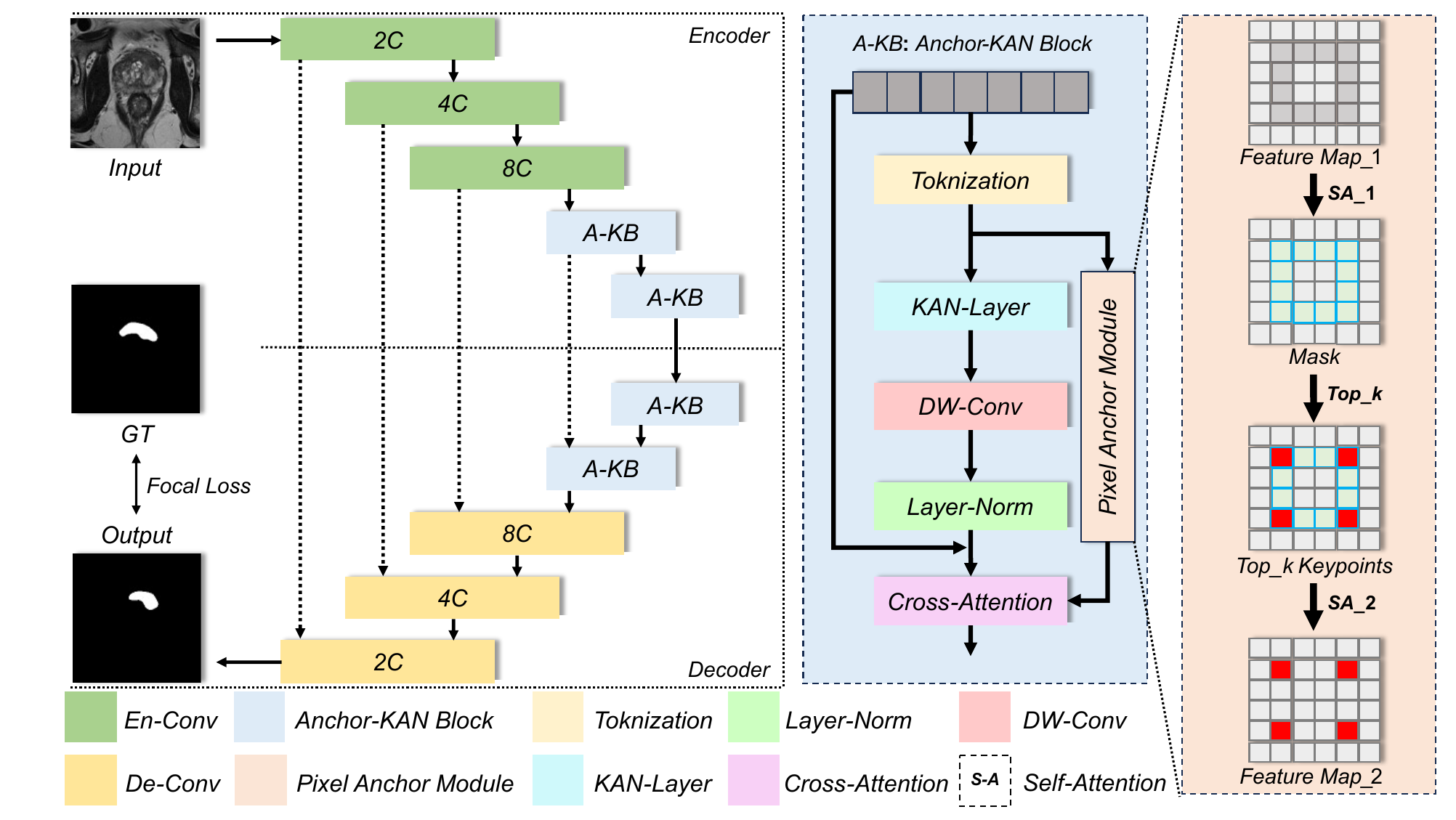}
		\caption{Overview of the proposed MyGO architecture, which is composed of the UKAN-baseline and the Anchor-KAN block. The Anchor-KAN block first tokenizes feature maps and employs the KAN-Layer to perform nonlinear transformations, enhancing representational capacity. Additionally, the Pixel Anchor module applies dual self-attention mechanisms and selects the Top\_$k$ keypoints to serve as pixel-wise anchors, thereby strengthening the model’s ability to capture global semantic structure.} 
		\label{fig_2}
	\end{figure*}
	
	\subsection{Self Attention}
	
	In 2017, the Transformer architecture was first introduced, employing self-attention to process entire input sequences \cite{vaswani2017attention}. Subsequent studies \cite{khan2022transformers, 9716741} demonstrated that Vision Transformer (ViT) and related approaches emerged around 2019 \cite{dosovitskiy2020image}. Furthermore, a self-attention mechanism designed specifically for image recognition was proposed in 2020 \cite{zhao2020exploring}, marking the integration of self-attention into computer vision tasks. By 2022, the combination of convolutional operations and self-attention enhanced the performance of models such as ACmix, which outperformed baseline methods in both image recognition and downstream tasks \cite{pan2022integration}. In 2024, the introduction of Beyond Self-Attention extended its application to medical image segmentation \cite{azad2024beyond}. Additionally, self-attention-based models have been employed for disease prediction tasks \cite{rahman2024enhancing}, demonstrating its substantial potential in medical imaging.
	
	\subsection{Self Position Point}
	
	Self Position Point primarily adapts to input shapes to select and localize key anchor points, enabling simultaneous pixel-space and semantic information processing while decoupling the attention mechanism. This enhances the model's representation capability. Such approaches have been widely applied in point cloud segmentation \cite{park2023self, zhang2025point}. Additionally, Self Position Point has contributed to dataset generation in related fields, such as a self-localization point dataset for vehicular networks\cite{7506097} proposed in 2016, and a vision-based drone self-localization dataset \cite{10376356}. However, research exploring the extension of Self Position Point to medical image segmentation remains limited. Applying this methodology to the detection and segmentation of prostate cancer lesions represents a pioneering direction in the field.
	
	\section{Proposed MyGO}
	
	\subsection{Overview}
	
	This paper addresses the limitations of existing segmentation models in accurately capturing lesion features, primarily due to semantic similarity between lesion and non-lesion tissues and the small size of lesion regions, both of which hinder reliable prostate cancer diagnosis. To overcome these challenges, we propose a Anchor KAN Block integrated into the U-KAN backbone. This block leverages the Pixel Anchor Module to assign pixel-level feature predictions within each homogeneous region to its respective pixel anchor. The Pixel Anchor Module incorporates a self-attention mechanism to assign region-specific weights, guiding anchor selection under the constraints of Pixel-wise Cross-entropy and Focal Loss. This facilitates the learning of semantic representations anchored to discriminative pixels within each region. Consequently, the proposed method consistently outperforms conventional CNN-based and Transformer-based architectures. The overall framework is depicted in \textbf{Figure 2}.  
	
	\subsection{U-KAN Baseline}
	
	In this work, we adopt U-KAN \cite{li2025u} as our baseline framework, which takes an MRI image as input, detects lesion regions, performs segmentation, and subsequently generates an output. The encoder and decoder components consist of convolutional layers, including 2C, 4C, and 8C, along with four Anchor-KAN blocks. Furthermore, each En-Conv and De-Conv layer is connected to its corresponding F-KB via skip connections to mitigate gradient issues associated with deep networks.
	
	Morever, Kolmogorov–Arnold Network (KAN)\cite{liu2024kan} is based on the Kolmogorov–Arnold representation theorem, which decomposes any continuous multivariate function 
	\( f(x_1, x_2, \dots, x_l) \) 
	into a nested composition of univariate continuous functions. This facilitates the construction of neural network architectures with enhanced interpretability. A traditional multilayer perceptron (MLP) is expressed as:
	
	\begin{equation} 
		\text{MLP}(x) = W_{L-1}\left( \sigma\left( W_{L-2}\left( \cdots W_1\left( \sigma\left( W_0(x) \right) \right) \cdots \right) \right) \right)
	\end{equation} 
	where, \( W_k \) are linear weights and \( \sigma \) is a fixed activation function. In contrast, KAN replaces this with:
	
	\begin{equation} 
		\text{KAN}(x) = \Phi_{L-1}\left( \Phi_{L-2}\left( \cdots \Phi_1\left( \Phi_0(x) \right) \cdots \right) \right)
	\end{equation} 
	\begin{equation} 
		\quad \Phi_l = \{ \varphi_{q,p}(x_{l,i}) \}
	\end{equation} 
	where, each \( \varphi_{q,p} \) is a learnable univariate function (e.g., B-spline), allowing network layers to directly operate on individual input components. This leads to strong expressive power with fewer parameters and a more transparent structural mechanism. The pseudo code of it is as \textbf{Algorithm 1}.

	\begin{algorithm}[!t]
		\caption{Kolmogorov--Arnold Network}
		\label{alg:kan_io}
		
		\textbf{Input:} Input vector $x_0 \in \mathbb{R}^{n_0}$, functional matrices $\{\Phi_l\}_{l=0}^{L}$ \\
		\textbf{Output:} Output vector $x_L \in \mathbb{R}^{n_L}$
		
		\begin{algorithmic}[1]
			\STATE Initialize $x \gets x_0$
			\FOR{$l = 0$ to $L-1$}
			\FOR{$j = 1$ to $n_{l+1}$}
			\STATE $x_{l+1,j} \gets 0$
			\FOR{$i = 1$ to $n_l$}
			\STATE $x_{l+1,j} \mathrel{+}= \varphi_{l,j,i}(x_{l,i})$
			\ENDFOR
			\ENDFOR
			\STATE $x \gets x_{l+1}$
			\ENDFOR
			\STATE \textbf{Return} $x_L \gets x$
		\end{algorithmic}
	\end{algorithm}

	\subsection{Anchor-KAN Block}
	
	We propsed a Anchor-KAN Block. The En-Conv layers tokenize pixel-wise features into tokens, which are refined by the KAN layer to enhance semantic discrimination, especially in distinguishing lesions from normal tissues. Depth-wise Convolution (DW-Conv) further captures spatial relations with high efficiency, followed by normalization for stable MRI performance. To enrich semantic learning, the Pixel Anchor Module aligns predictions within each region to its Anchor, while skip connections between (En/De)-Conv and Anchor-KAN blocks preserving essential spatial information and mitigating gradient vanishing issues in deep networks.

	\subsection{Pixel Anchor Module}
	
	In conventional Transformer-based models, attention computation typically requires processing n points, leading to high computational complexity. We propose a novel Pixel Anchor Module, which restructures the attention mechanism by first initializing a central point and subsequently propagating the correlation among central points to all feature points. This approach enables the module to establish global feature connectivity through central point correlations.
	
	The module first initializes central points, generating $Feature Map_1$, and establishes inter-central point connections to ensure global feature propagation. Initially, self-attention is employed to compute an attention map, followed by a Top\_$k$ selection operation to extract the Top\_$k$ most relevant points. These selected $k$ central points are then interconnected via attention mechanisms, producing $Feature Map_2$. Finally, cross-attention is utilized to propagate global feature representations from the central points to the entire feature space.
	Where the point set generated by the Top\_$k$ selection operation is illustrated in \textbf{Figure 3}.

	
	In this process, a self-attention mechanism is employed to derive attention weights for focusing on distinct regions. Regions with higher weights contribute more significantly to the segmentation outcome and are thus considered more critical for overall performance. Consequently, the pixel point with the highest attention weight is selected as an anchor. Pixel-wise Cross-entropy Loss and Focal Loss are subsequently applied to constrain the learning at this anchor location. The segmentation results are then utilized to iteratively optimize the anchor selection strategy.
	
	\begin{algorithm}[!htpb]
		\caption{Forward Propagation of Pixel Anchor Module}
		\label{alg:pixel_Anchor_module_refined}
		
		\textbf{Input:} Feature map $FM_1 \in \mathbb{R}^{C \times H \times W}$ \\
		\textbf{Output:} Refined feature map $FM_2 \in \mathbb{R}^{C' \times H \times W}$
		
		\begin{algorithmic}[1]
			\STATE Select central points via attention: 
			
			$C \gets \texttt{AttentionSelect}(FM_1)$
			
			\STATE Compute self-attention among central points: 
			
			$A \gets \texttt{SelfAttention}(C)$
			
			\STATE Select top-$k$ key points based on $A$: 
			
			$C_k \gets \texttt{TopK}(A,\,k = 0.25 \times H \times W)$
			
			\STATE Aggregate features via refined attention: 
			
			$FM_2 \gets \texttt{SelfAttention}(C_k)$
			
			\STATE \textbf{Return} $FM_2$
		\end{algorithmic}
	\end{algorithm}

	\subsection{Loss Function}
	
	Our Loss Function consists of two components. The first component serves as our baseline, incorporating the loss function from UKAN's foundational work, which employs \textbf{pixel-wise cross-entropy loss} as its optimization criterion. The second component integrates the \textbf{focal loss} \cite{Lin_2017_ICCV}, which we introduce to enhance segmentation performance.
	
	\subsubsection{\textbf{Pixel-wise Cross-entropy Loss}}
	
	Pixel-wise cross-entropy loss is a specialized variant of the conventional cross-entropy loss, tailored for image segmentation tasks. It calculates the cross-entropy loss individually for each pixel and subsequently averages the loss across all pixels within the image. The formulation of Pixel-wise Cross-entropy Loss is defined as follows:
	\begin{equation} 
		\mathcal{L_{CE}} = -\frac{1}{N} \sum_{i=1}^{N} \sum_{c=1}^{C} y_{i,c} \log(\hat{y}_{i,c})
	\end{equation}
	where, $N$ denotes the total number of pixels in the image. $y_{i,c}$represents the ground-truth label for pixel $i$ belonging to class $c$
	, encoded in a one-hot format. $\hat{y}_{i,c}$ corresponds to the predicted probability that pixel $i$ belongs to class $c$, typically obtained from a softmax function.
	
	\subsubsection{\textbf{Focal Loss}}
	
	In prostate cancer lesion segmentation, lesions are small and often resemble normal tissue, leading to class imbalance. To address this, we apply Focal Loss to help the model focus on challenging samples and improve segmentation performance.
	The formulation of Focal Loss is defined as follows:
	\begin{equation}
		FL(p_t) = - \alpha_t (1 - p_t)^\gamma \log(p_t)
	\end{equation}
	where, $p_t$ represents the predicted probability of the correct class by the model. If the true class is $y = 1$, then $p_t = p$; if the true class is $y = 0$, then $p_t = 1 – p$. The term $log(p_t)$ corresponds to the standard cross-entropy loss, which is used to measure the confidence of the prediction. The modulation factor $(1 - p_t)^\gamma$ plays a crucial role, for easily classified samples with high $p_t$, $(1 - p_t)^\gamma$ approaches zero, reducing their loss contribution; whereas for hard-to-classify samples with low $p_t$, $(1 - p_t)^\gamma$ remains close to one, resulting in a higher loss weight. The parameter $\alpha_t$ serves as a weighting factor, ensuring a balance between positive and negative samples.
	
	\subsubsection{\textbf{Total Loss}}
	
	The total loss is the sum of the original baseline loss, the pixel-level cross entropy loss and the focal loss we introduced, and its formula is expressed as
	\begin{equation} 
		L_{total} = \mathcal{L_{CE}}(y_{i,c}, \hat{y}_{i,c}) + Focal(y, p_t)
	\end{equation}
	where, $y$ and $y_{i,c}$ both represent the true label at the pixel level, and $\hat{y}_{i,c}$ and $p_t$ represent pixel-level prediction probabilities. 
	
	\begin{figure*}[!t]
		\centering
		\includegraphics[width=0.9725\textwidth]{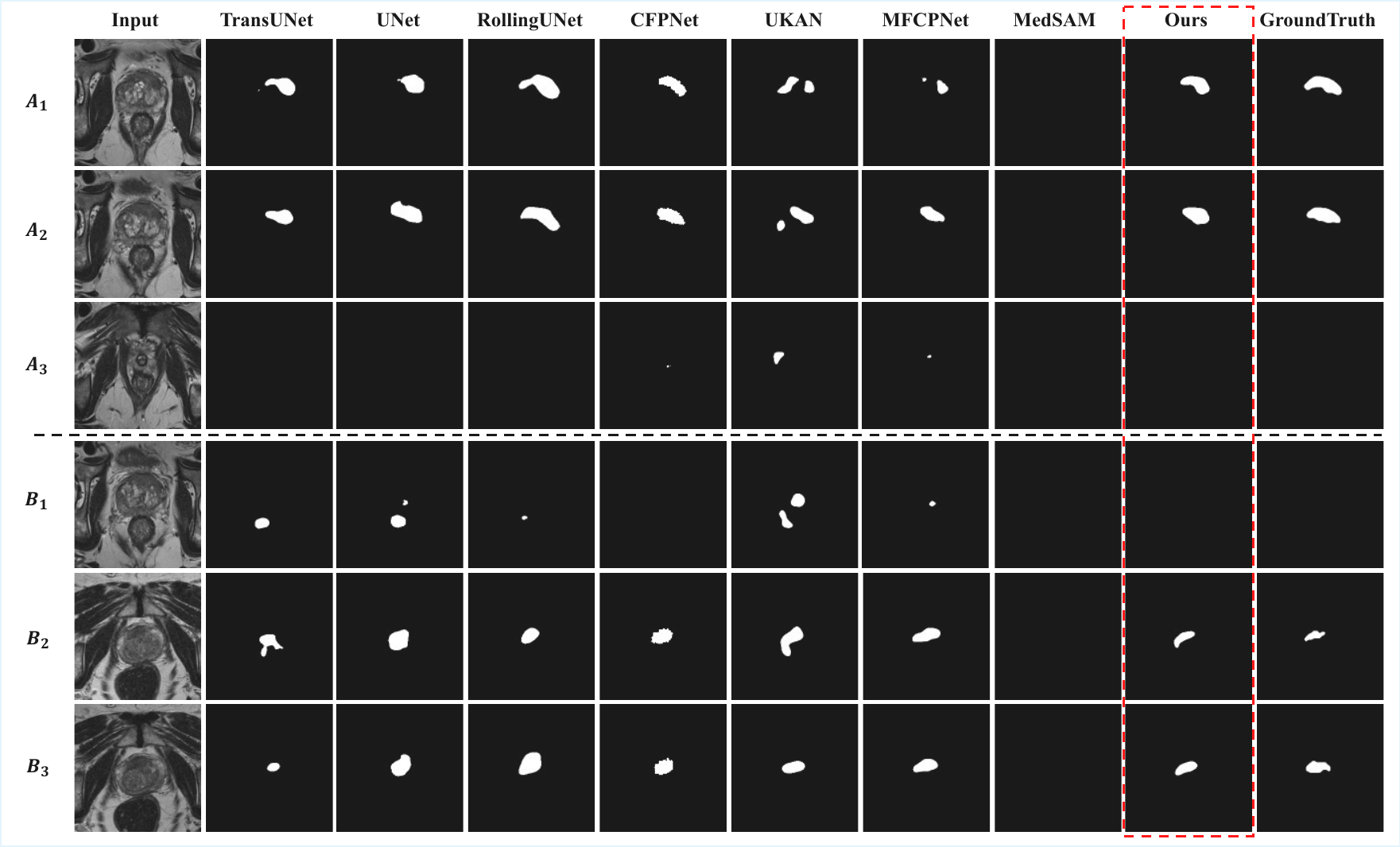}
		\caption{We conduct a qualitative comparison of segmentation performance on the public PICAI dataset against existing SOTA methods. Among them, A and B are two different cases, $A_1$ $\to$ $A_3$ and $B_1$ $\to$ $B_3$ are slices at different locations in the prostate.Notably, the segmentation results produced by our approach exhibit texture and region shape that are more consistent with the ground truth in both form and size.} 
		\label{fig_3}
	\end{figure*}
	
	\section{Experiment}
	
	\begin{table*}[!t]
		\centering
		\renewcommand{\arraystretch}{1.0} 
		\setlength{\tabcolsep}{10pt} 
		\begin{tabular}{l|c|c|c|c|c|c}
			
			\hline
			Methods & Venue &IoU(\%)$\uparrow$ & Dice(\%)$\uparrow$ &Specificity(\%)$\uparrow$  & F1 Score(\%)$\uparrow$ &  FPR(\%) $\downarrow$\\ 
			
			\hline
			\hline
			
			UNet
			& MICCAI'15 & 24.00 & 35.64 & 99.57 & 21.02& 0.43  \\ 
			
			TransUNet
			& arxiv'21 & 29.67 & 39.35 & 99.69 & 13.11 & 0.31 \\
			
			UCTransNet
			& AAAI'22 & 0.59 & 1.16 & - & 1.25 & - \\ 
			
			CFPNet
			& CBM'23 & 40.01 & 51.21 & 99.79 & \underline{21.74}& 0.21  \\
			
			RollingUNet  
			& AAAI'24 & 43.31 & 54.65 & 99.64 & 21.11 & 0.36 \\ 
			
			MFCPNet  
			& BSPC'24 & 51.69 & 61.36 & 99.76 & 17.36 & 0.24 \\ 
			
			UKAN          
			& AAAI'25 & \underline{66.82} & \underline{72.94}   & 99.66 & \textbf{25.37}& 0.34\\ 
			\hline
			\hline
			
			Ours                      
			& - & \textbf{69.73} & \textbf{74.32} & \textbf{99.87} & 19.02 & \textbf{0.13}\\ 
			\hline		
		\end{tabular}
		\caption{Compare With other SOTA Methods on PI-CAI Dataset}
		\label{NCI}
	\end{table*}
	
	\subsection{PI-CAI Dataset}
	
	The Prostate Imaging: Cancer AI (PI-CAI) dataset is a large-scale MRI dataset specifically designed for prostate cancer detection \cite{saha2024artificial}. It is jointly provided by multiple medical institutions in the Netherlands and Norway. The primary objective of this dataset is to facilitate the advancement of artificial intelligence applications in prostate cancer diagnosis.
	The dataset comprises 9,000–11,000 prostate MRI scans, collected from four medical centers across the Netherlands and Norway. The imaging modalities include T2-weighted (T2W) sequences, diffusion-weighted imaging (DWI), and apparent diffusion coefficient (ADC) maps. Among these, the Public Training and Development Dataset (1,500 cases) is made available for public research and development of AI models.
	
	\subsection{Evaluation Metrics}
	
	To ensure a fair and comprehensive comparison between our method and existing SOTA methods, we have selected four evaluation metrics: IoU(\%), Dice Score(\%), Specificity(\%), F1 Score(\%) and False Positive Rate (FPR(\%)).
	
	\subsection{Experimental Sets}
	
	We select U-KAN as the baseline for our model and configure the training parameters as follows. 
	First, we choose the PI-CAI dataset and set the batch size to 16. 
	The learning rate is initialized at 0.0001, and Adam is employed as the optimization algorithm. 
	Additionally, we utilize a cosine annealing learning rate scheduler and 
	set the minimum learning rate to 0.00001 to enhance the training performance of the model.
	
	The formula for the cosine annealing learning rate scheduler is as follows:
	\begin{equation}
		\eta_t = \eta_{\min} + \frac{1}{2} (\eta_{\max} - \eta_{\min}) \left(1 + \cos\left(\frac{T_{\text{cur}}}{T_{\text{max}}} \pi\right)\right)
	\end{equation}
	where, $\eta_t$ represents the current learning rate, with $\eta_min$ and $\eta_max$ denoting its minimum and maximum values, respectively. Additionally, $T_{cur}$ refers to the ongoing training step, while $T_{max}$ indicates the total number of training steps.
	
	We set the backbone network to train for a total of 400 epochs. The dataset consists of 34 groups, with the first 32 groups used for training and the remaining 2 groups reserved for testing. Each group contains between 18 to 26 MRI images. These images are from T2-Weighted Images. Our experimental environment and equipment information are as follows: PyTorch: 1.10.1; Python: 3.7 (Ubuntu 22.04); CUDA: 11.1; GPU: RTX 4060ti (16GB).
	
	\subsection{Compare with Sota Methods}
	
	We compare our approach with existing SOTA methods on the PI-CAI dataset. Specifically, we evaluate the following methods,
	U-Net \cite{ronneberger2015u},
	TransUNet \cite{chen2021transunet},
	CFP-Net \cite{lou2023cfpnet},
	UCtransNet \cite{wang2022uctransnet},
	Rolling Unet \cite{liu2024rolling},
	MFCPNet \cite{hou2025mfcpnet},
	U-KAN \cite{li2025u}, 
	and the comparative results are illustrated in \textbf{Table 1}, which is mainly used to solve the problem of unbalanced sample distribution, where our method demonstrates the closest resemblance to the Ground Truth. 
	
	Under identical experimental settings, our model achieves higher val IoU (\%) and Dice (\%) scores in test results compared to U-KAN, as shown in \textbf{Figure 4}.
	
	\subsection{Compare with Large Segment Model}
	
	We further evaluated large segmentation models, including MedSAM \cite{ma2024segment} and SAM \cite{kirillov2023segment}, using both checkpoint-based inference and demo testing. Results indicate that when directly applying the trained checkpoints, the models failed to produce meaningful responses over lesion regions. This suggests that semantic confusion arising from highly homogeneous semantic information continues to hinder lesion identification. For details, refer to the MedSAM results in \textbf{Figure 3}.
	In the demo evaluation of SAM, visual segmentation outcomes on test data revealed that while SAM successfully delineates anatomical contours within the prostate on lesion-containing MRI images, it fails to recognize lesion areas \textbf{Figure 5}. These findings underscore that, despite their strong general segmentation capabilities, neither SAM nor MedSAM currently incorporate mechanisms tailored to address challenges posed by semantic homogeneity in lesion localization.
	
	\begin{figure*}[!t]
		\centering
		\includegraphics[width=1.0\textwidth]{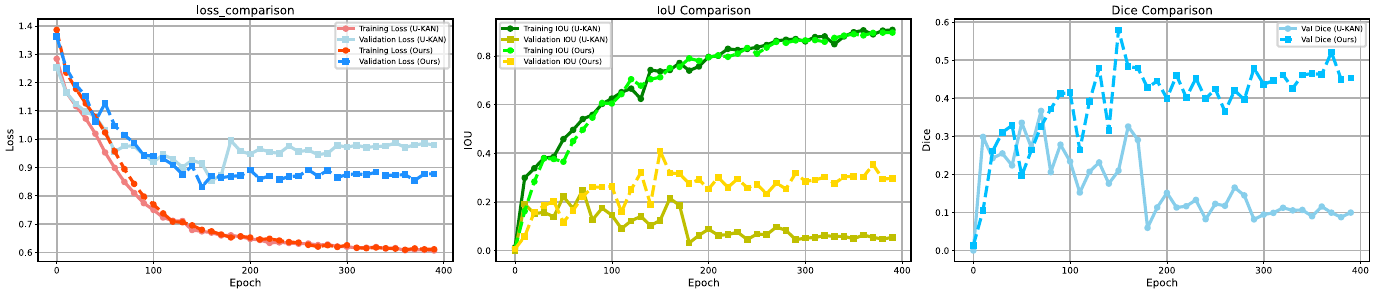}
		\caption{Experimental Figure: In PI-CAI dataset, we compared with the best performing method at present, and the comparison results are shown in the figure; as the figure shows that ours' training performance is better than U-KAN. We mainly compared the three indicators of IoU, Dice and Loss coefficient.} 
		\label{fig_5}
	\end{figure*}
	
	\begin{figure}[!t]
		\centering
		\includegraphics[width=\columnwidth]{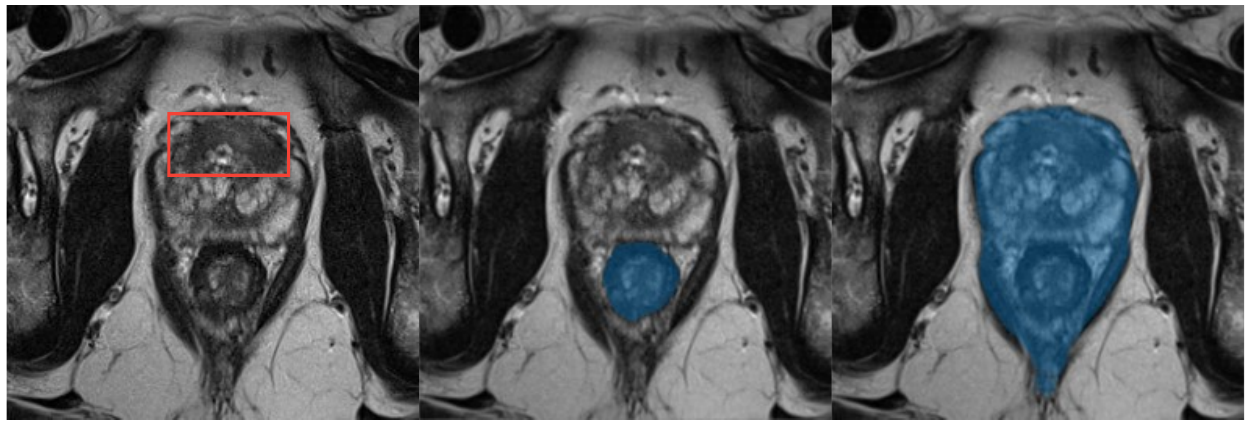} 
		\caption{The figure presents the visualization results from demo testing of the SAM model. The test indicates that SAM is highly sensitive to anatomical boundaries of the prostate region; however, it fails to respond to lesion areas. Notably, the region enclosed by the red bounding box corresponds to the lesion.}
		\label{fig_4}
	\end{figure}
	
	\subsection{Ablation Study}
	
	\textbf{Table 2} summarizes the performance impact of various modules in MyGO using IoU (\%), Dice (\%), and Specificity (\%) metrics; where, we note Specificity(\%) as Spec.(\%) in \textbf{Table 2}. We conduct ablation studies by progressively integrating or modifying components on top of the U-KAN baseline. These components include Focal Loss (FL), Pixel Anchor Module (PAM), and self-attention (SA) of PAM at different Top\_$k$ operation stages. Results indicate that while individual modules may cause performance degradation when applied in isolation, their joint activation consistently enhances segmentation performance. The optimal configuration (denoted as Ours) enables all subcomponents, demonstrating their complementary effectiveness.
	
	\subsubsection{Effect of Pixel Anchor Module}
	We extend the baseline model by integrating the proposed Pixel Anchor Module. The test results exhibit a slight performance drop, which we attribute to the loss of fine-grained details caused by sample imbalance. To address this issue, we introduce Focal Loss for further constraint. Additionally, we conduct an independent evaluation of the Focal Loss component.
	
	\subsubsection{Effect of Focal Loss}
	Applying Focal Loss directly to the original baseline model leads to a modest decline in segmentation performance, suggesting that it cannot independently improve model capability. This finding implies that sample imbalance only becomes significant when the Pixel Anchor Module is activated.
	
	\subsubsection{Effect of Top\_$k$ \& Self Attention}
	
	\textbf{SA\_1 before Top\_$k$:} Under Focal Loss constraint, we perform ablation studies on the internal components of the Pixel Anchor Module. Both attention variants utilize the same self-attention mechanism, and to differentiate their placement around the Top-$k$ operation, we denote the pre-Top-$k$ variant as $SA_1$ and the post-Top-$k$ variant as $SA_2$. Experimental results with $SA_1$ placement show minimal performance change, indicating that without downstream attention refinement, the Top-$k$ feature map lacks global anchor connectivity and offers limited segmentation benefit.
	
	\textbf{Top\_$k$ before SA\_2:} Building upon the $SA_1$ configuration, we reposition the attention operation to follow Top-$k$, forming $SA_2$. Test results show improvements with IoU and specificity increasing by 2.16\% and 0.2\% respectively. This demonstrates the pivotal role of post-Top-$k$ attention in enhancing semantic representation of selected anchors.
	
	\subsubsection{Overall Performance}
	The complete MyGO model fuses both SA$_1$ and SA$_2$ around the Top-$k$ module to enable bidirectional attention flow. This design achieves the best overall performance: 69.73\% IoU, 74.32\% Dice, and 99.87\% specificity. The joint configuration facilitates cross-layer feature alignment and minimizes semantic degradation, effectively bridging shallow appearance cues with high-level semantics. These results validate the effectiveness of our proposed module.
	
	\begin{table}[!t]
		\centering
		\renewcommand{\arraystretch}{1.025} 
		\setlength{\tabcolsep}{0.5pt} 
		\begin{tabular}{ l | c | c | c }
			
			\hline
			Modules & IoU(\%) & Dice(\%) & Spec.(\%)  \\ 
			
			\hline
			Baseline  & 66.82          & \underline{72.94}          & 99.66          \\ 
			\hline
			\hline
			
			Focal Loss (FL)    & 64.58   & 68.96  & 99.81        \\
			
			PAM (SA\_1 \& Top\_$k$ \& SA\_2)          & 65.22   & 69.46   & 99.85   \\
			
			\hline
			\hline
			
			FL \& PAM (SA\_1 \& Top\_$k$)  & 65.46  & 70.45   & 99.83        \\ 
			
			FL \& PAM (Top\_$k$ \& SA\_2)  & \underline{68.98}  & 72.56   & \underline{99.86}         \\ 
			
			\hline 
			\hline
			Ours (all)  & \textbf{69.73} & \textbf{74.32}  &\textbf{99.87}  \\ 
			\hline 
			
		\end{tabular}
		\caption{Ablation Study}
		\label{ABLATION}
	\end{table}

	\section{Conclusion}
	This paper presents an innovative Pixel Anchor Module designed to address the semantic confusion problem in prostate cancer MRI segmentation. The proposed module leverages a minimal set of feature anchors to capture and comprehend global features, thereby enhancing nonlinear modeling capability and improving lesion region identification accuracy. Furthermore, the Top\_$k$ selection mechanism based on self-attention refines feature anchor recognition, leading to optimal performance on the PI-CAI dataset and significantly improving prostate cancer lesion segmentation effectiveness.
	Experimental results demonstrate that our method outperforms current state-of-the-art approaches.
	In future work, we will further investigate the impact of semantic confusion on lesion quantification and clinical risk stratification.
	
	
	
	\bibliography{aaai2026.bib}
\end{document}